\documentclass[%
reprint,
 amsmath,amssymb,
 aps,
]{revtex4-1}

\usepackage{graphicx}
\usepackage{dcolumn}
\usepackage{bm}
\usepackage{hyperref}

\usepackage{subfigure, color}

\def\D{\displaystyle}

\begin{document}

\preprint{APS/123-QED}

\title{Seeded Ising Model and  Statistical Natures of
Human Iris Templates}

\author{Song-Hwa Kwon}
 \email{Corresponding author. skwon@catholic.ac.kr}
 \thanks{This work was supported by the Catholic University of Korea, Research Fund, 2016.}
\affiliation{%
Department of Mathematics, Catholic University of Korea,
43 Jibong-ro, Bucheon-si, Gyeonggi-do 14662, Republic of Korea 
}%

\author{Hyeong In Choi}
\affiliation{%
Department of Mathematics, Seoul National University, 1 Gwanak-ro, Gwanak-gu, Seoul 08826, Republic of Korea 
}%


\author{Sung Jin Lee}
\affiliation{
 Department of Mathematics, Daejin University,
 1007 Hoguk-ro, Pocheon-si, Gyeonggi-do 11159, Republic of Korea 
}%

\author{Nam-Sook Wee}
\affiliation{
 Division of Smart Management Engineering, Hansung University, 116 Samseongyo-ro 16-gil, Seongbuk-gu, Seoul 02876, Republic of Korea
}%


\date{\today}

\begin{abstract}
We propose a variant of Ising model, called the Seeded Ising Model, to model probabilistic nature of human iris templates. This model is an Ising model in which the values at certain lattice points are held fixed throughout Ising model evolution.
Using this we show how to reconstruct the full iris template from partial information, and we show that about 1/6 of the given template is needed to recover almost all information content of the original one in the sense that the resulting Hamming distance is well within the range to assert correctly the identity of the subject.
This leads us to propose  the concept of effective statistical degree of freedom of iris templates and show it is about 1/6 of the total number of bits. In particular, for a template of $2048$ bits, its effective statistical degree of freedom is about $342$ bits,
which coincides very well with the degree of freedom computed by the completely different method proposed by Daugman.
\end{abstract}

\pacs{Valid PACS appear here}
\maketitle


\section{Introduction}
Human iris texture exhibits very intricate, even random or chaotic patterns.
It is known that no two human beings, even identical twins, have the same iris pattern. Exploiting this fact, Daugman invented an iris recognition method which is widely used as a means of identifying human  individuals \cite{Daugman1993, Daugman2004}.

Daugman's method, roughly put, creates a two-dimensional binary array called template or iris code from the annular iris region using Gabor transform; and the templates are matched using the Hamming distance.

\begin{figure}[b]
\includegraphics[width=0.45\textwidth]{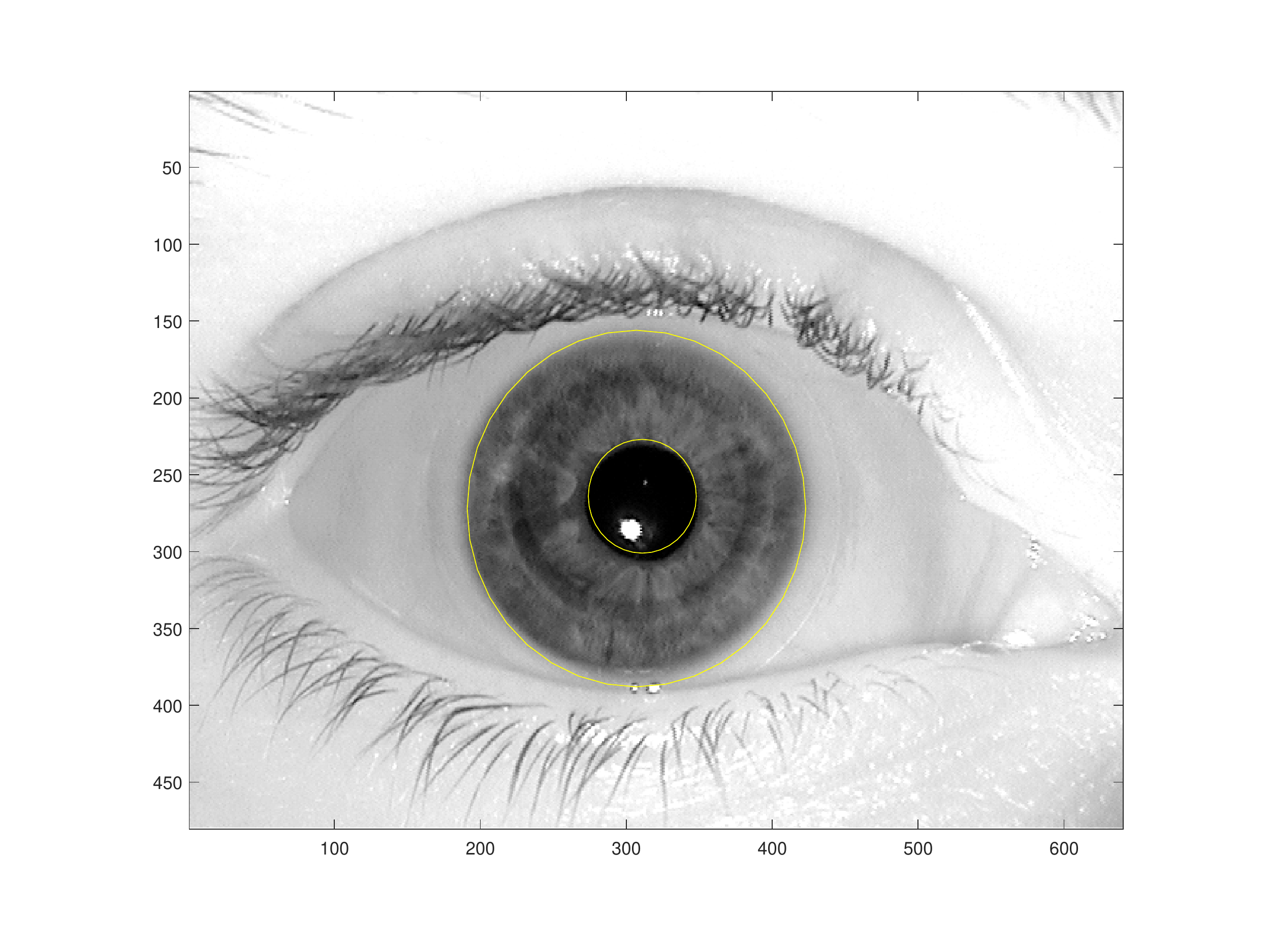}
\caption{\label{fig:eyeimage} A sample of eye image (from ICE2005)}
\end{figure}

\begin{figure*}[t]
{%
\setlength{\fboxsep}{0pt}%
\setlength{\fboxrule}{0.5pt}%
\fbox{\includegraphics[width=0.95\textwidth]{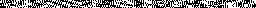}}%
}%
\caption{\label{fig:iristemplate1} Iris template from the image in Fig.~\ref{fig:eyeimage}. }
\end{figure*}


\begin{figure}[t]
    \centering
    \begin{subfigure}
        \centering
        {%
        \setlength{\fboxsep}{0pt}%
        \setlength{\fboxrule}{0.5pt}%
        \fbox{\includegraphics[width=0.45\textwidth]{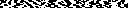}}%
        }%
    \end{subfigure}%

    \begin{subfigure}
        \centering
        {%
        \setlength{\fboxsep}{0pt}%
        \setlength{\fboxrule}{0.5pt}%
        \fbox{\includegraphics[width=0.45\textwidth]{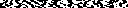}}%
        }%
    \end{subfigure}%

\caption{\label{fig:iristemplate2} Real (top) and imaginary (bottom) parts of the iris template in Fig.~\ref{fig:iristemplate1}. }
\end{figure}

Figure \ref{fig:iristemplate1} and \ref{fig:iristemplate2} show a template in which the black pixels are the bits whose binary values are 0 and the white ones with binary value 1. One can easily observe that 0s and 1s tend to cluster together in somewhat randomly alternating fashion.
The clusters represent folding patterns of the 3-dimensional shapes which are the result of stochastic or possibly chaotic development in
utero whose detailed morphogenesis depends on initial conditions in the embryonic mesoderm.


%

In this paper, we study a probabilistic model which we hope will elucidate this phenomenon of random mixing with clustering.
The model we propose is what we call the {\em Seeded Ising Model}. It is an Ising model \cite{Cipra1987} in which bits in certain locations are held fixed throughout Ising model dynamic evolution.
This way, our Seeded Ising Model reconstructs a template from a fraction of the information of the whole template.

\begin{figure*}[t]
{%
\setlength{\fboxsep}{0pt}%
\setlength{\fboxrule}{0.5pt}%
\fbox{\includegraphics[width=0.95\textwidth]{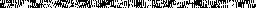}}%
}%
\caption{\label{fig:rec_iristemplate1} Reconstructed iris template }
\end{figure*}

It turns out that this model has a few remarkable properties.
First of all, the reconstructed template exhibits similar visual quality
when compared qualitatively with the original one. (See Figure \ref{fig:rec_iristemplate1}.) However, more remarkable is the fact that the reconstructed templates actually match the original ones rather quite accurately when computed with Hamming distance.
This means that the reconstructed templates retain quite a bit of information content of the original ones as far as iris recognition is concerned.

It is better to elaborate a bit more on seeds.
First of all, it is true that the  morphogenetic initial conditions relating to the embryonic development of mesoderm and ectoderm are important factors for the full biological development of iris, which again is responsible for the fractal or chaotic patterns of iris.
However, it is very hard to  pin down biologically these initial conditions.
One can only guess they must be somehow random in nature. Similarly, in our model seeds are chosen randomly, which is meant to be a mathematical abstraction of the random nature of biological initial conditions.

The clustering phenomenon of iris template indicates that not all binary bits can be independent. Then a question arises as to what is its degree of freedom. Daugman looked at this question by examining the impostor distribution. By approximating it with a binomial distribution, he claims that the degree of freedom must be about 12.16\% (249 bits  out of 2048 bits) \cite{Daugman2004}.
In this paper we use somewhat different template generation and matching algorithm \cite{Lee13}. The impostor distribution gotten by the algorithm we use is somewhat narrower, which implies that the degree of freedom computed {\`a} la Daugman is about 17.19\% (352 bits  out of 2048 bits). (For details, see Fig. \ref{fig:ImpDistFit2}.)

We also look at this problem of degree of freedom from a completely different angle. We examine the genuine, not impostor, matching and find that our Seeded Ising Model needs 342 bits as seeds out of 2048 bits  in order to recover quite faithfully  the information content of the original template.
It translates into about 16.70\% (342 bits  out of 2048 bits). It is remarkable to notice that the degrees of freedom gotten by two completely different approaches coincide so well.

The degree of freedom computed this way is only statistical in nature and furthermore, it only relates to the information content from the view point of iris recognition. Because of this reason, we call it the {\em Effective Statistical Degree Of Freedom}.

As far as we are aware, this kind of Seeded Ising Model was not so far studied anywhere. However, it exhibits some remarkable physical property resembling re-normalization phenomenon. We hope to be able to publish this finding elsewhere.

First, in Section \ref{sec:model}, we describe the proposed Seeded Ising Model, and present a method of reconstructing iris templates from partial information. It is a sampling  method based on the Metropolis algorithm adapted to our model.
We then examine how well our model reflects the nature of human iris templates by conducting, in Section \ref{sec:experiments}, a series of statistical experiments based on the proposed model.  Subsequently we discuss the concept of effective statistical degree of freedom, and finally give concluding remarks in Section \ref{sec:conclusion}.

%
%
%
%



\section{Seeded Ising Model}\label{sec:model}

%
%

A real part or imaginary part of the human iris template is modeled by a binary random field $x$ on an $m \times n$ regular lattice.
As we said above, each bit of iris template is binary with value  $0$ or $1,$
but when it comes to the Ising model presentation, we use the convention that each bit has value $-1$ or $1,$ instead.
This convention is purely for the sake of simplicity of notation of Ising model.
So according to this convention, $0$ in iris template is replaced with $-1$ for the Ising model, and vice versa.
With this notational convention, which should be understood in context throughout this paper,
the space of all iris templates is denoted by

$$
\mathcal{T} = \{ x | x_{i,j} \in \{-1,1\} \hbox{ for } 1 \le i \le m,\,\, 1 \le j \le n \}.
$$

In the following description, we use the univariate indexing to denote the position in an $m \times n$ regular lattice by utilizing a mapping, for example, $(i,j) \mapsto k=i + (j-1)\times m.$ With the univariate indexing scheme, the space of iris templates can be simply written by $\mathcal{T} = \{x | x_k \in \{-1,1\} \hbox{ for } 1 \le k \le m n \}.$

For a given subset $I \subseteq \{1, 2, \cdots, m n \},$ a map $s : I \to \{-1, 1\}$ is regarded as a `partial template data' specifying the value of the template at positions in $I.$ So, the set of templates which have the same partial template data as $s$ is denoted by $\mathcal{T}(s) = \{ x | x_k = s(k) \hbox{ for each } k \in I \}.$

All the templates in $\mathcal{T}(s)$ have the same value at positions of $s,$ and  the values at other positions of a template in $\mathcal{T}(s)$
may be regarded as having grown from the fixed `seed' $s.$
Our way of looking at this ``growth'' is in  fact picking the most `likely' sample from $\mathcal{T}(s)$ under some probability distribution.

For a given seed $s,$ we model the probability distribution  $P(x)$ on the space of $\mathcal{T}(s)$ by
$$
P(x) = \frac{1}{Z}\exp\left( \sum_{i \sim j} J_{i,j} x_i  x_j \right),
$$
where $i \sim j$ means the positions $i$ and $j$ are adjacent to each other, thus, the sum is done over all adjacent positions, $J_{i,j}$ is a constant parameter for the adjacent positions $i$ and $j$, and $Z$ is the partition function given by
$$Z = \sum_{x \in \mathcal{T}(s)} \exp\left( \sum_{i \sim j} J_{i,j} x_i  x_j \right).$$

Since we use different $J_{i,j}$ depending on whether $i$ and $j$ are horizontally or vertically adjacent to each other,
we say the relation $i \sim_v j$ means that two positions $i$ and $j$ are vertically adjacent to each other, and the relation $i \sim_h j$ that two positions $i$ and $j$ are horizontally adjacent to each other with circular-end conditions employed for each row of a template. Note that we think of the first column and the last column in a regular lattice are adjacent to each other with circular-end conditions employed.
In this paper, we set $J_{i,j} = J_v$ when two positions $i$ and $j$ are adjacent vertically, and $J_{i,j} = J_h$ when two positions $i$ and $j$ are adjacent horizontally. With these conventions, $P(x)$ can be written as
$$
P(x) = \frac{1}{Z}\exp\left( J_{v} \sum_{i \sim_v j}  x_i  x_j  + J_{h} \sum_{i \sim_h j}  x_i  x_j \right).
$$
We call this probabilistic model of the space of templates $\mathcal{T}(s)$ with seed $s$ the \emph{Seeded Ising Model.}

\subsection{Sampling via Metropolis Algorithm}
For a given seed $s: I \to \{-1, 1\}$, we sample templates in $\mathcal{T}(s)$ according to the distribution defined by $P(x)$ via the Metropolis algorithm \cite{newman1999}. First note that, under the Seeded Ising Model, the probability $P(x)$ is proportional to
\begin{eqnarray*}
  &&\exp\left( J_{v} \sum_{ i \sim_{v} j} x_i x_j +  J_{h} \sum_{ i \sim_{h} j} x_i x_j \right) \\
  &&= \exp\left( J_v ( mn - n - 2d_x^v )+J_h(mn-2d_x^h)\right),
\end{eqnarray*}
 where $d_x^v$ denotes the number of disagreeing vertical edges in template $x$ and $d_x^h$ denotes the number of disagreeing horizontal edges. Thus, $P(x)$ is also proportional to the un-normalized probability $\pi(x) = \exp(-2J_v d_x^v - 2J_h d_x^h).$

Let a template $x \in \mathcal{T}(s)$ be represented by a vector
$$
x = (x_1, x_2, \cdots, x_{k-1}, x_k, x_{k+1}, \cdots, x_{mn}).
$$
Then, the Metropolis algorithm modified for our context would have the following steps:
\begin{enumerate}
  \item Start with an initial template $x \in \mathcal{T}(s).$
  \item Select randomly a \emph{non-seed index} $$k \in \{1,2,\cdots, mn \} \setminus I.$$
  \item Propose a new template $x'$ as
  $$
  x'= (x_1, x_2, \cdots, x_{k-1}, -x_k, x_{k+1}, \cdots, x_{mn}).
  $$

  \item Define the proposal probability of $t_1 \to t_2,$ moving from a template $t_1$ to a template $t_2$ by
  $$
  Q(t_1 \to t_2) = \left\{
    \begin{array}{lll}
      \D \frac{1}{m n - |I|}, & &\hbox{if } t_1, t_2 \hbox{ differ at exactly}\\& &\hbox{one non-seed index.}\\
      \\
           0, & &\hbox{otherwise.}
    \end{array}
  \right.
  $$
Then, accept $x'$ with probability $\mathcal{A}(x\to x'),$
$$
\mathcal{A}(x\to x') = \min\left[1, \frac{\pi(x') Q(x'\to x)}{\pi(x) Q(x\to x')} \right],
$$
where $\D \pi(x) = \exp(-2J_v d_x^v - 2J_h d_x^h)$ is the un-normalized probability of $x.$ Since $x, x'$ differ in exactly one non-seed index $k$ by the construction of $x'$ from $x,$ $Q(x\to x') = Q(x'\to x) > 0,$ and thus we get
$$
\mathcal{A}(x\to x') = \min\left[1, \frac{\pi(x')}{\pi(x)} \right].
$$

The ratio in $\mathcal{A}(x\to x')$ is
\begin{eqnarray*}
&&\exp(-2J_v(d_{x'}^v - d_{x}^v)-2J_h (d_{x'}^h - d_{x}^h) ) \\
&&= \exp(2J_v(d_{x, k}^v - a_{x, k}^v) +2J_h(d_{x, k}^h - a_{x, k}^h)),
\end{eqnarray*}
where $d_{x, k}^v$ is the number of disagreeing vertical edges between the index $k$ and its vertically adjacent indices in template $x,$ $a_{x, k}^v$ is the number of agreeing vertical edges for $x$ at $k.$ $d_{x, k}^h$ and $a_{x, k}^h$ are defined similarly.

\item Generate a uniform random number $u \in (0,1)$ and accept $x'$ as the current template if $u < \mathcal{A}(x\to x').$ Otherwise, keep $x$ as the current template and go to Step 2.
\end{enumerate}

\subsection{Reconstruction of Iris Templates}

By the Metropolis algorithm for the Seeded Ising Model, we can sample a template $x \in \mathcal{T}(s)$ for a given seed $s$ according to the distribution defined by $P(x).$ However, since the space $\mathcal{T}(s)$ is so huge, it is highly unlikely that one sampled template can serve as a representative template in $\mathcal{T}(s).$ For this reason, we produce a reconstructed template by using the idea of bagging ({\bf b}ootstrap {\bf agg}regat{\bf ing}) \cite{Breiman1996A} in machine learning. The procedure of obtaining a reconstructed template is as follows: Let $t^0$ be an initial template for Metropolis algorithm, and $t^n$ be the template obtained after $n$ iterations by the Metropolis algorithm. Then, for predetermined positive integers $L$ and $n_j$'s for $j = 1, 2, \cdots, L$ with the condition that $0 \le n_1 < n_2 < \cdots < n_L,$ we define the reconstructed template $r^{\mathbf{n}}$ whose value $r^{\mathbf{n}}_i$ at position $i$ is given by the following:
\begin{equation}\label{eq:reconstruction}
r^{\mathbf{n}}_i =
\left\{
  \begin{array}{ll}
    1, & \hbox{ if } \D \sum_{j=1}^{L} t^{n_j}_i  \ge 0 \\
    -1, & \hbox{ if } \D \sum_{j=1}^{L} t^{n_j}_i  < 0 ,
  \end{array}
\right.
\end{equation}
where $\mathbf{n}$ denotes the vector of $\mathbf{n}=(n_1, \cdots, n_L).$

Fig. \ref{fig:rec_iristemplate1} shows an example of template $r^{\mathbf{n}}$ reconstructed from a seed $s$ that is extracted from a full human iris template in Fig. \ref{fig:iristemplate1}. When two templates in Fig. \ref{fig:iristemplate1} and in Fig. \ref{fig:rec_iristemplate1} are compared, the visual qualities look quite similar even though the reconstructed template $r^{\mathbf{n}}$ is generated by utilizing only the partial information in Fig.  \ref{fig:iristemplate1}. This may be taken as a supporting evidence that our Seeded Ising Model explains reasonably well the statistical nature of human iris templates, at least visually.

%
%

\section{Statistical Experiments}\label{sec:experiments}

To examine how good the proposed Seeded Ising Model is, we performed statistical experiments. In the experiments, we used the algorithm developed by \citet{Lee13} for iris template generation and matching. The size of iris template generated by this algorithm is $8 \times 256$ which is of the same size as the template in Fig.~\ref{fig:iristemplate1}. Since the template consists of real part and imaginary part, the size of real (or imaginary) part of templates as in Fig.~\ref{fig:iristemplate2}
is $8\times 128.$

The dataset used in this paper for statistical experiments consists of reasonably good images selected from ICE2005 Dataset which was used for Iris Challenge Evaluation 2005 \cite{Phillips08}. Table~\ref{tab:ICE2005Dataset} shows basic statistics of ICE2005 Dataset and Table~\ref{tab:GoodICE2005Dataset} shows basic statistics of the dataset used in this paper.

\begin{table}[ht]
\caption{\label{tab:ICE2005Dataset}ICE2005 Dataset used (with error corrected\footnote{\cite{Phillips08}})}
\begin{ruledtabular}
  \begin{tabular}{ccc}
  Position & \# of Images & \# of Subjects\\\hline
  Left & 1527 & 119 \\\hline
  Right & 1426 & 124 \\\hline
  Total & 2953 & 132 \\
  \end{tabular}
  \end{ruledtabular}
\end{table}

\begin{table}[ht]
\caption{\label{tab:GoodICE2005Dataset} Dataset used in this paper}
\begin{ruledtabular}
  \begin{tabular}{ccc}
  Position & \# of Images & \# of Subjects  \\\hline
  Right & 948 & 120  \\\hline
  & \# of Genuine matchings & \# of Impostor matchings \\\hline
  &  5,953 & 442,925 \\
  \end{tabular}
  \end{ruledtabular}
\end{table}

\begin{figure}[t]
\includegraphics[width=0.5\textwidth]{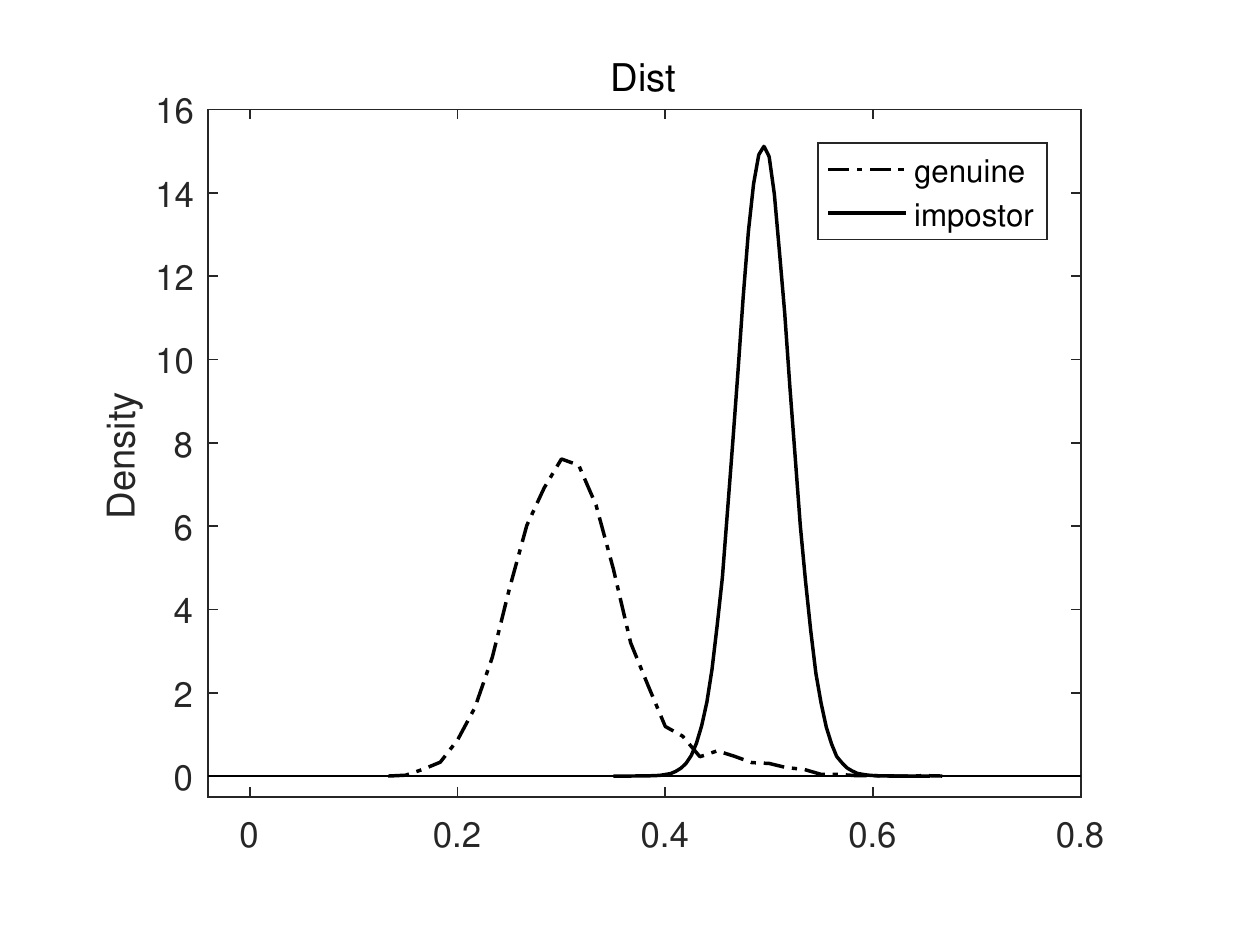} 
\caption{\label{fig:GenImpDistNoRotation} Genuine/Impostor Distribution of Hamming distance without rotation (5,953 genuine matchings, 442,925 impostor matchings) }
\end{figure}

\begin{figure}[t]
\includegraphics[width=0.5\textwidth]{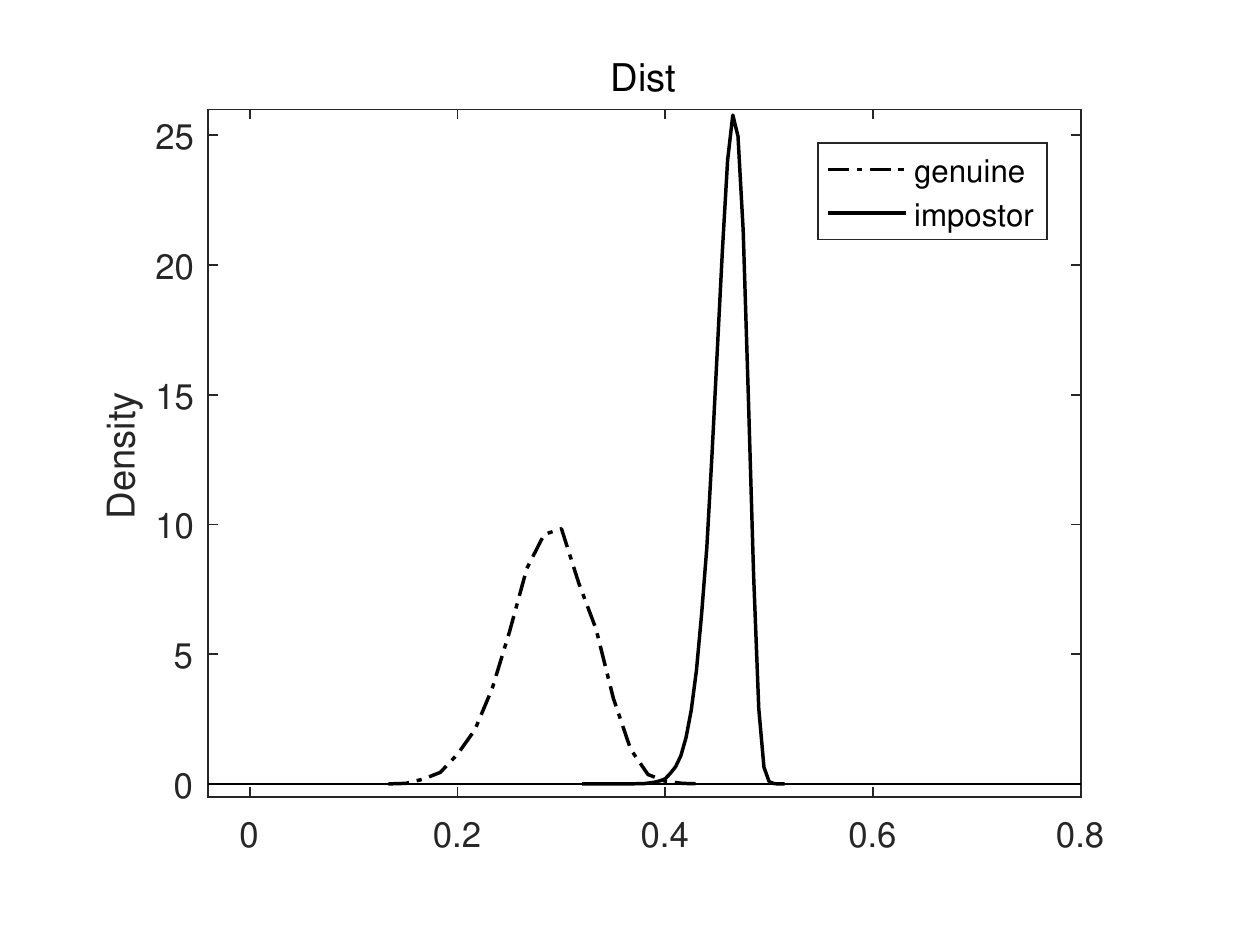} 
\caption{\label{fig:GenImpDistWithRotation} Genuine/Impostor Distribution of Hamming distance with rotation (5,953 genuine matchings, 442,925 impostor matchings)}
\end{figure}


Fig.~\ref{fig:GenImpDistNoRotation} shows the matching results of the dataset used in this paper when no rotation is considered in matching, and Fig~\ref{fig:GenImpDistWithRotation} shows the matching results of the same dataset with rotation applied. From the distribution, we may observe that most of genuine matching (matching two iris templates from the same person) has Hamming distance less than $0.4,$ and most of impostor matching (matching two iris templates from two other persons) has Hamming distance more than $0.4.$ So, roughly speaking, the best threshold that distinguishes whether a given pair of iris templates comes from the same person or not would be around $0.4.$

\subsection{The parameter $J$}\label{sec:J}

To verify our Seeded Ising Model explains well the statistical nature of iris templates, we first need to find the parameter $J=(J_v, J_h).$ To determine the best parameter, we first performed statistical experiments for a range of parameter values. For the experiment, we first fixed one iris template $x = (x^R, x^I)$ where $x^R$ and $x^I$ denote the real and imaginary part of template $x,$ respectively, and for each
$$
J=(J_v, J_h) \in \{(0.1*i,0.1*j) |\,\, 1 \le i,\, j \le 10\, \},
$$
$100$ templates are reconstructed using random seeds $s=(s^R, s^I),$ where $s^R$ and $s^I$ denote the seed for the real and imaginary part of templates, respectively, and matched with the original template $x.$ To obtain each reconstructed template, we first randomly chose an index set $I \subseteq \{1, 2, \cdots, mn \},$ where $mn$ is the size of real or imaginary part of template $x,$ which is $1024(=8\times128)$ in our case. In this experiment, we chose $I$ so that $|I| = 256,$ i.e., the ratio of information in seed $s$ to that in $x$ is $1/4.$ After obtaining the index set $I,$ the seed $s^R:I\to \{-1,1\}$ is naturally given from template $x^R$ by the relation that $s^R(k) = x^R_k$ for $k \in I.$ The seed $s^I$ for the imaginary part of template is also determined similarly from $x^I.$ Once the seed $s^R$ is determined, the real part of a reconstructed template is computed by Equation (\ref{eq:reconstruction}) with ${\mathbf{n}} = 10^4(1,2,\cdots,100).$ The imaginary part of a reconstructed template is also generated from the seed $s^I$ similarly, and these two parts are combined to produce the final reconstructed template.

\begin{figure}[t]
\includegraphics[width=0.48\textwidth]{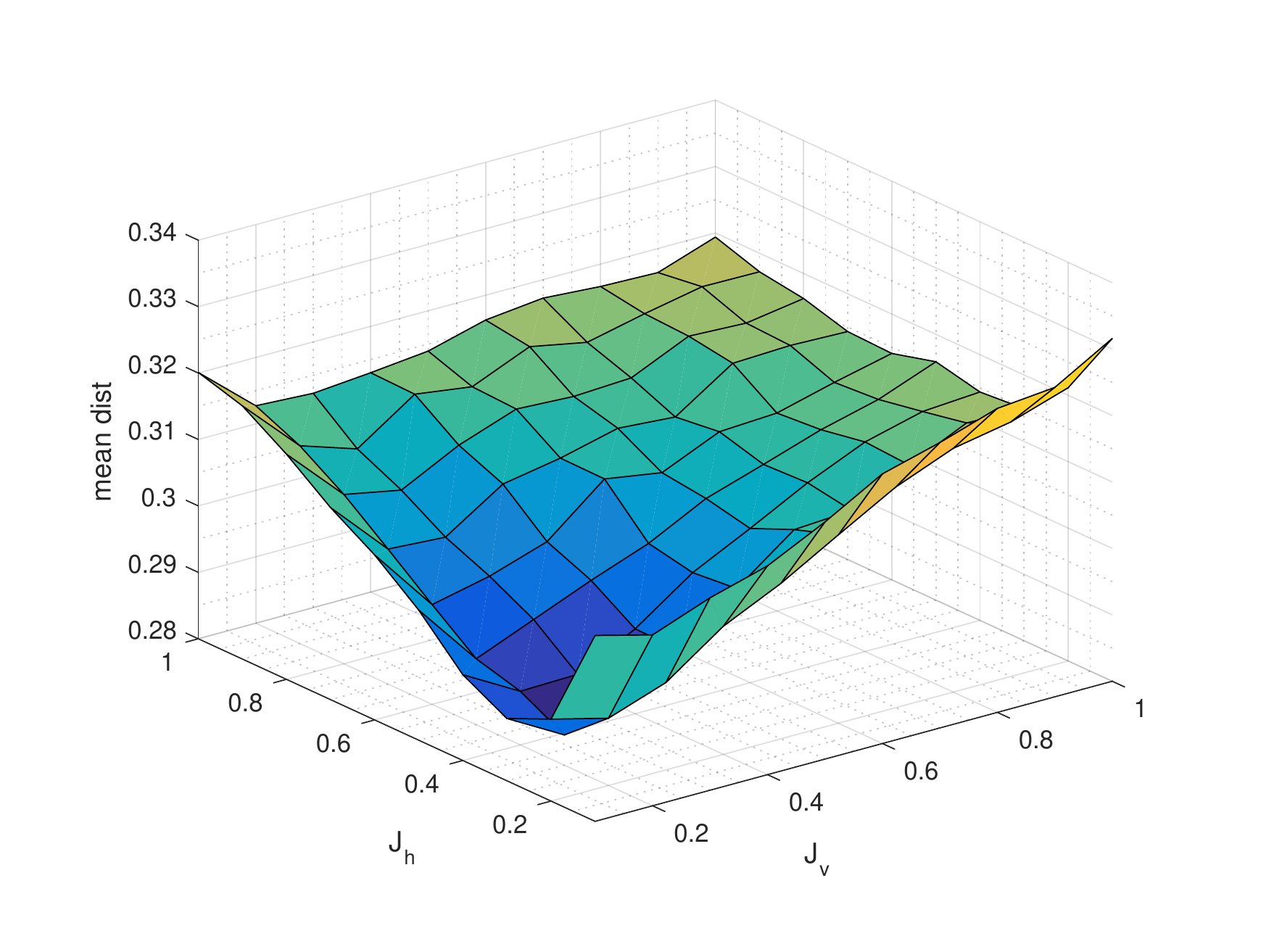}
\caption{\label{fig:avgTplDistMeans} Minimum mean distance (0.2846) at $J=(0.2, 0.3).$}
\end{figure}

The average of 100 Hamming distances for each value of $J=(J_v, J_h)$ is shown in Figure~\ref{fig:avgTplDistMeans}. The minimum of $0.2846$ of mean Hamming distance is obtained when $J=(J_v, J_h) = (0.2, 0.3).$
In the subsequent experiments, we use this values for the parameter $J$ unless mentioned otherwise.

\subsection{Hamming Distances of Reconstructed Templates}

To compare the original iris template and the reconstructed template, we selected $100$ images from ICE2005 Dataset, and generated $100$ reconstructed templates for each iris template and measured Hamming distance between the original template and the reconstructed template. For reconstruction, we used the same procedure as described in Section \ref{sec:J} with several different values of seed size, i.e., $1/5, 1/6, 1/7$ of template size.

In what follows, an initial template is the one randomly selected from  $\mathcal{T}(s)$ as in Step 1 of the Metropolis algorithm.  In practice, initial template is made by randomly assigning values at non-seed positions, while using the fixed values of $s$ at the positions of seed $s.$

\begin{figure}[t]
\includegraphics[width=0.48\textwidth]{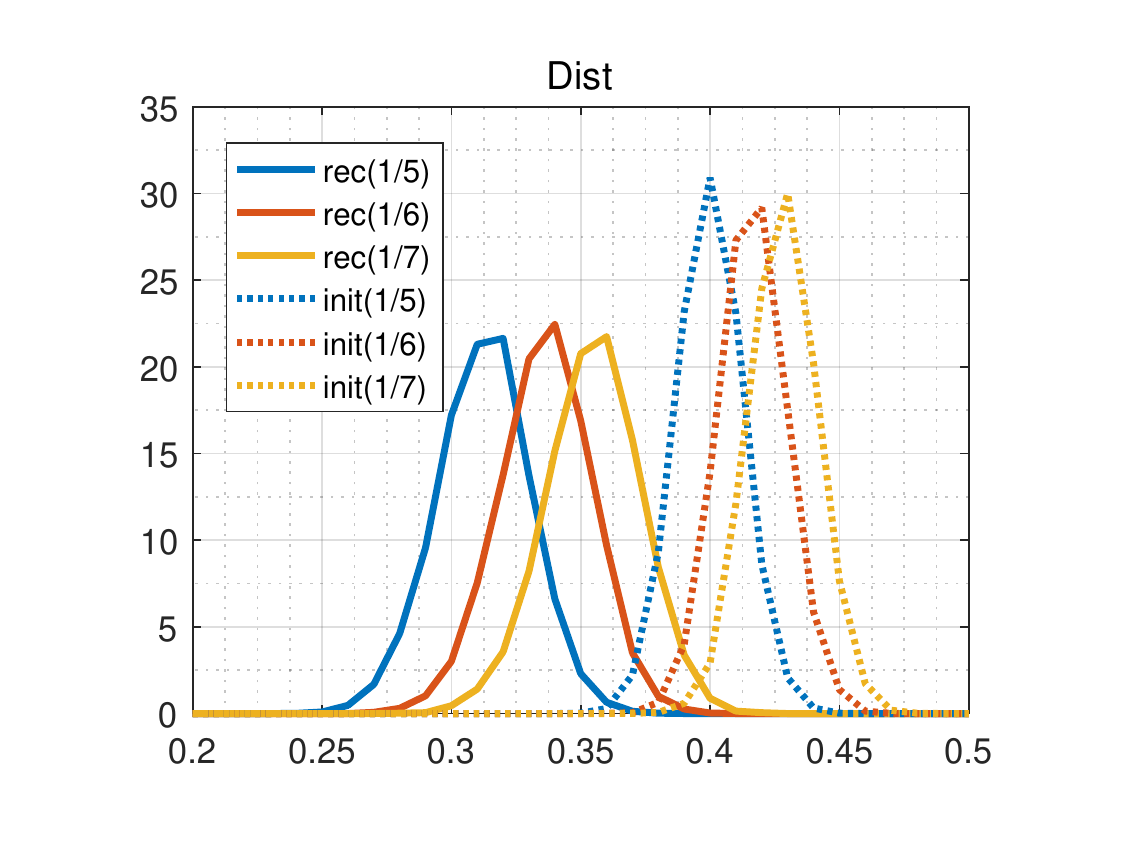} 
\caption{\label{fig:recTplDistHist}}
\end{figure}

\begin{table}[t]
\caption{\label{tab:ReconstructedTplDistAll}Means and Standard Deviations of distances}
  \begin{ruledtabular}
    \begin{tabular}{ccccc}
            & \multicolumn{2}{c}{Initial Template} & \multicolumn{2}{c}{Reconstructed Template} \\\cline{2-5}
      size of seeds      & mean & std & mean & std \\\hline
      $1/5$ & 0.3998 & 0.0126 & 0.3123 & 0.0179 \\
      $1/6$ & 0.4164 & 0.0128 & 0.3362 & 0.0179 \\
      $1/7$ & 0.4281 & 0.0130 & 0.3547 & 0.0182
    \end{tabular}
  \end{ruledtabular}
\end{table}



Figure~\ref{fig:recTplDistHist} shows the distributions of Hamming distances. In Figure~\ref{fig:recTplDistHist}, the dotted curves represent the distributions of Hamming distances between the original templates and the initial templates in the reconstruction procedure, and the solid curves represent the distributions of Hamming distances between the original templates and the reconstructed templates given by Equation (\ref{eq:reconstruction}).
The numbers $1/5, 1/6, 1/7$ in the legend represent the proportion of the seed size in the full template.
Table~\ref{tab:ReconstructedTplDistAll} shows the mean distance and standard deviation for each case. Compared to the mean Hamming distance of initial templates, the mean Hamming distance of reconstructed templates is significantly small. For instance, in the case of  $1/6$ seed size, the mean distance of $0.3362$ is more than $3.5\sigma$ away from the threshold value of $0.4,$ which is
a relatively good threshold determining whether two iris templates come from the same person or not as discussed in the beginning of this section.
Even in the case of $1/7$ seed size, the mean distance is more than $2.4\sigma$ away from the threshold.
This implies that the reconstructed template is statistically similar to, or essentially have the same information content as, the original iris template.
In other words, the reconstructed template can be safely deemed to have come from the same person to whom the original template belongs.
This, we believe, is a good evidence that the proposed Seeded Ising Model is a reasonably good model for probabilistic model of iris templates.

\subsection{Effective statistical degree of freedom} \label{DoG}
Once the Seeded Ising Model has been shown as a good model for iris templates, the next question would be how big the seed size is in order to reliably reconstruct human iris templates. For that, we use the cumulative distribution function of Hamming distance.

\begin{figure}[b]
\includegraphics[width=0.48\textwidth]{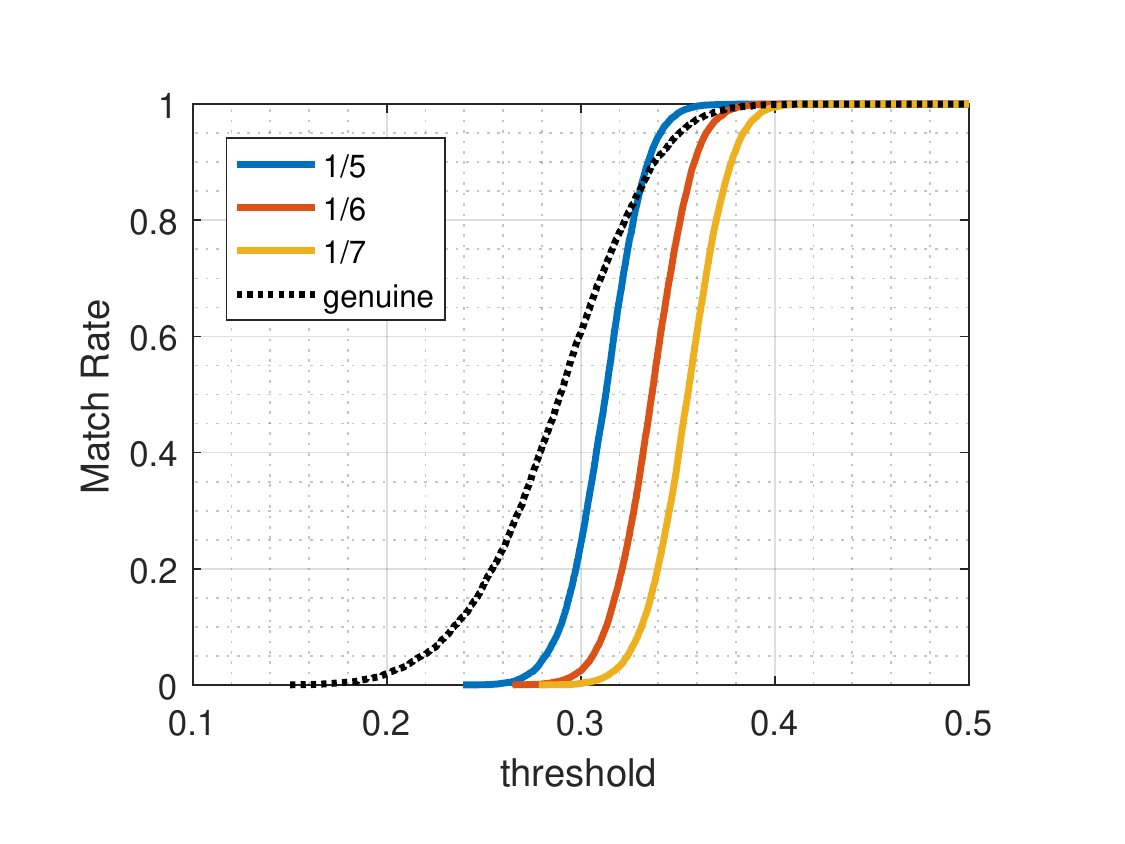} 
\caption{\label{fig:avgTplMatchRate}}
\end{figure}

For a given collection of $M$ matching distances $(d_1, d_2, \cdots, d_M),$ match rate $m(d)$ is defined as the ratio of the number of matchings with Hamming distance less than or equal to $d,$

$$
m(d) = \frac{|\{ i  | d_i \le d \}|}{M}.
$$

Figure~\ref{fig:avgTplMatchRate} shows match rate curves for several different collections of matching distances. The dotted curve represents the matching rate for genuine matchings of 948 images.
(See Table~\ref{tab:GoodICE2005Dataset} for the details.) The solid curves in blue, red, yellow represent the matching rates of Hamming distances between the original templates and reconstructed templates from the seeds of size $1/5, 1/6, 1/7,$ respectively.

\begin{figure}[t]
\includegraphics[width=0.48\textwidth]{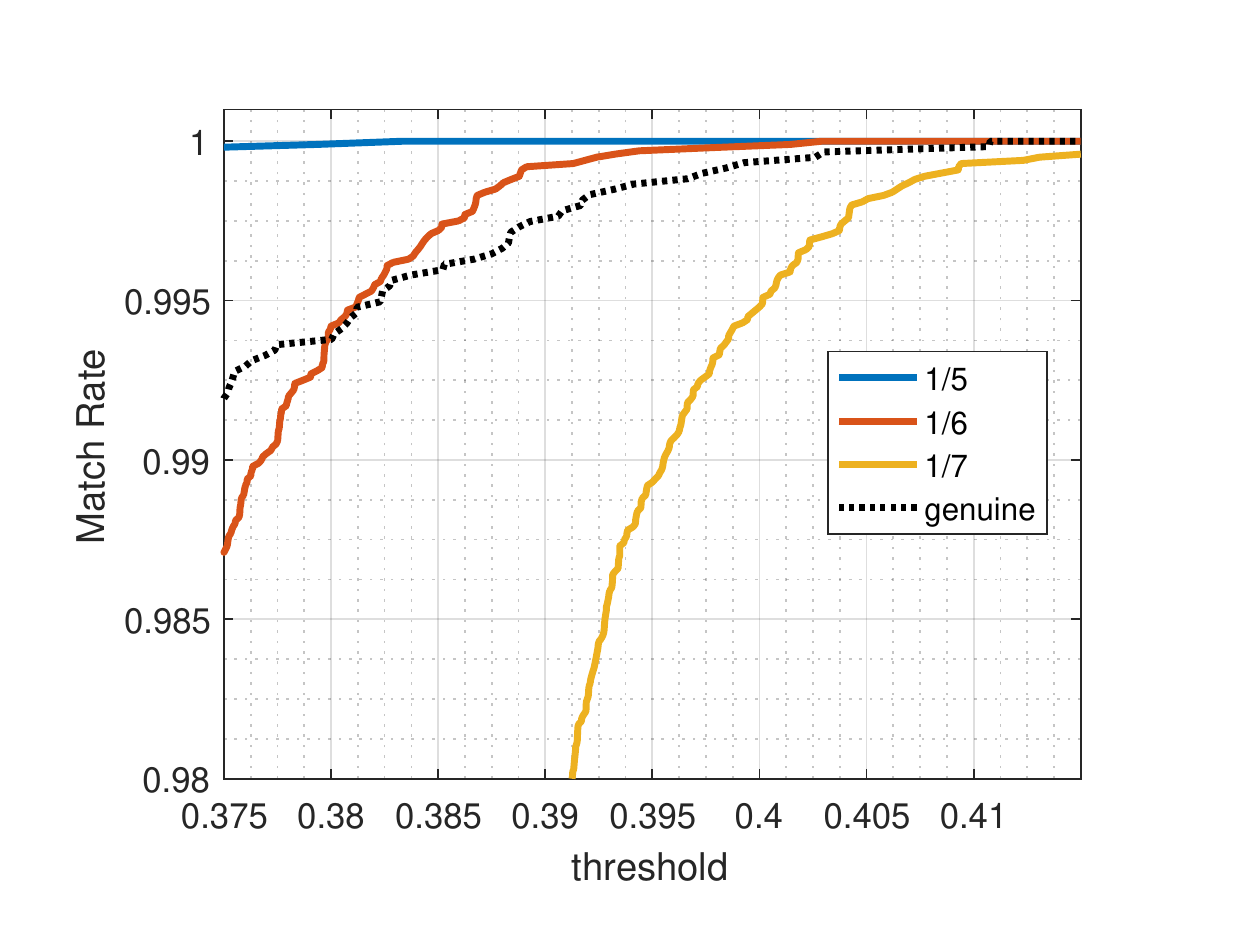} 
\caption{\label{fig:avgTplMatchRateMagnified}}
\end{figure}



Figure~\ref{fig:avgTplMatchRateMagnified} is the zoomed figure of Figure~\ref{fig:avgTplMatchRate} around Hamming distance $0.395.$ From these figures, we may say the matching rate of the reconstructed templates from the seeds of size $1/6$ is comparable to, with slightly better performance, the matching rate of the genuine matchings of real iris templates. This analysis gives approximate size of seeds for the real iris templates. Roughly speaking, only about $1/6 \approx 16.7\%$ of template information is needed to account for the information content of the whole iris templates through our Seeded Ising Model. From this, we may conclude that the proposed Seeded Ising Model needs $342$ bits as seeds out of $2048$ bits in order to recover quite faithfully  the information content of the original template. It translate into about $16.70\%$ ($342$ bits out of $2048$ bits).

As we alluded in the beginning, the clustering phenomenon of iris templates indicates that not all binary bits can be independent, and the above result suggests that the degree of freedom, the size of seeds, is about $16.7\%.$  The degree of freedom computed this way is only statistical in nature and furthermore, it only relates to the information content from the view point of iris recognition. Because of this reason, we call it the {\em Effective Statistical Degree Of Freedom}.

\begin{figure}[t]
\includegraphics[width=0.5\textwidth]{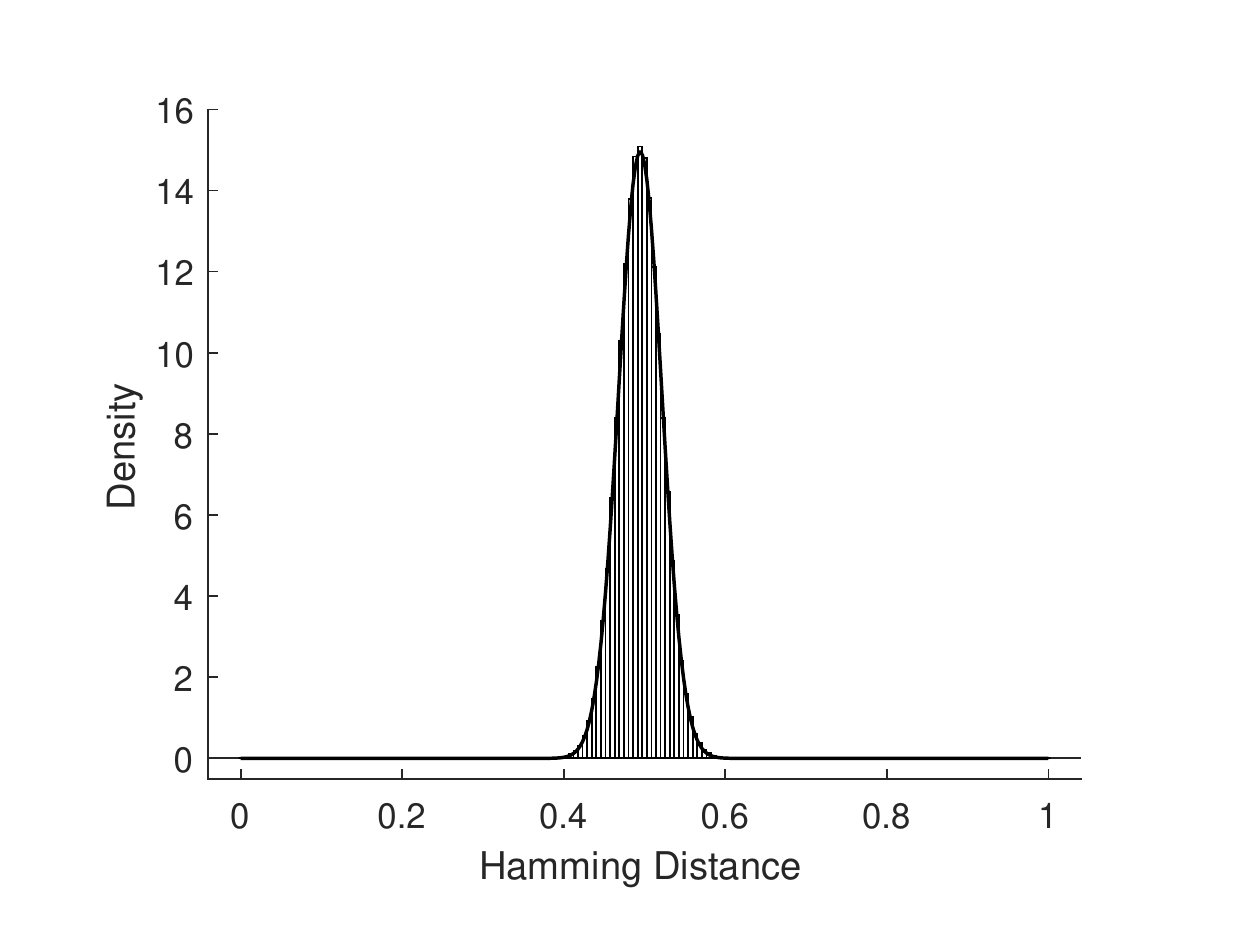}
\caption{\label{fig:ImpDistFit2} Distribution of Hamming Distances from $442,925$ impostor comparisons. The histogram is fitted by the binomial distribution with $p=0.4947$ and $N=352$ degrees-of-freedom in solid curve (\ref{eqn:binomial}). }
\end{figure}

Our result has a parallel counterpart coming from different angle.
Concerning the question of what is the degree of freedom of iris templates, Daugman examined the impostor distribution. (Ours is through the genuine distribution.) By approximating it with a binomial distribution given by $f(x)$,

\begin{equation}\label{eqn:binomial}
f(x) = \frac{N!}{m!(N-m)!}p^m (1-p)^{N-m},
\end{equation}
where $x = m/N$ is the outcome fraction of $N$ Bernoulli trials with probability of success $p,$ he claims that the degree of freedom must be about 12.16\% (249 bits  out of 2048 bits) \cite{Daugman2004}.

To replicate Daugman's finding in our context, we need to exercise some caution.
First, the algorithms we use for template generation and matching \cite{Lee13} is different from the one he uses, and we are also using different image dataset in this paper.
The impostor distribution plot in Figure~\ref{fig:ImpDistFit2} looks somewhat different from that in \cite{Daugman2004} in that our impostor distribution is somewhat narrower.
When we calculate, using our algorithm, the binomial distribution approximation as defined by $f(x)$ in Equation (\ref{eqn:binomial}), we got $p = 0.4947$ and $N=352.$
This implies that the degree of freedom calculated from the impostor distribution as suggested by Daugman is about 17.19\% (352 bits  out of 2048 bits), which is very similar to $16.7\%,$ our result obtained above through the genuine matching via template reconstruction from seeds.

It is remarkable to notice that the our definition of degree of freedom coincides so well with that of Daugman, although
the two approached to degree of freedom are completely different.

\section{Conclusion}\label{sec:conclusion}

In this paper, we proposed the Seeded Ising Model, a probabilistic model of human iris templates. Inspired by the biological processes how human iris texture patterns develop, we devised a probabilistic model by introducing `seeds' that is a mathematical abstraction of initial conditions in the embryonic
mesoderm and ectoderm from which the chaotic iris texture develops. Also, we found the best parameter value of $J$ for the proposed Seeded Ising Model from the iris recognition point of view. We also provided statistical evidences that the proposed model is a reasonably good model capable of explaining the probabilistic nature of human iris templates. In fact, it turns out that artificially generated iris templates based on the proposed model with the best parameter value of $J$ share many probabilistic natures of real human iris templates in that (1) The reconstructed templates are visually similar to real iris templates including clustering phenomenon; (2) The reconstructed templates from the partial template information are much more closer to the original iris templates in terms of Hamming distance.

We also found approximate size of seeds for the real iris templates. Roughly speaking, our Seeded Ising Model implies that
only about $1/6 \approx 16.7\%$ of template information is needed to account for the information content of the whole iris templates. Based on this finding. we propose the concept of {\em Effective Statistical Degree Of Freedom} of human iris templates. Surprisingly, this estimated value coincides very well with the degree of freedom computed by the completely different method proposed by Daugman.

Many evidences presented in this paper suggest that we may conclude the proposed model reflects well the real nature of human iris templates.



%
%
%
%
%
%
%

\bibliography{SIM2017}

\end{document}